\documentclass[a4paper,reqno]{amsart}
\usepackage[text={5.2in,8in},centering]{geometry}

\usepackage{amssymb}
\usepackage{enumerate}
\usepackage{graphicx}

\usepackage{xspace}

\usepackage[usenames,dvipsnames,svgnames,table]{xcolor}
\definecolor{citegreen}{rgb}{0.00,0.70,0.30}
\usepackage[colorlinks=true,
            linkcolor=blue,
            urlcolor=blue,
            citecolor=blue]{hyperref}

\usepackage{amsrefs}

\definecolor{redd}{rgb}{0.95,0.2,0.2}
\definecolor{gris}{rgb}{0.9,0.9,0.9}
\definecolor{greenn}{rgb}{0.1,0.6,0.2}

\definecolor{cmgray}{rgb}{0.7,0.7,0.7}

\definecolor{cmblue}{rgb}{0.2,0.5,0.8}

\definecolor{blue}{rgb}{0,0,1.0}
\def\tblue#1{\textcolor{blue}{#1}}

\numberwithin{equation}{section}
\theoremstyle{plain}
\newtheorem{theorem}{Theorem}

\newtheorem{lemma}{Lemma}[section]
\theoremstyle{remark}
\newtheorem{remark}{Remark}[section]

\newtheorem*{quest*}{Question}
\newtheorem*{remark*}{Remark}
\theoremstyle{remark}
\newtheorem{example}{Example}[section]
\theoremstyle{definition}
\newtheorem{definition}{Definition}[section]
\newtheorem*{definition*}{Definition}
\newtheorem*{notation*}{Notation}
\newtheorem*{notations*}{Notations}
\providecommand{\B}{\mathbf}

\providecommand{\C}{\mathcal}
\providecommand{\D}{\mathbb}

\providecommand{\R}{\mathrm}

\newcommand{\ee}{\mathrm{e}}

\DeclareMathOperator{\dist}{dist}
\DeclareMathOperator{\card}{card}

\DeclareMathOperator{\supp}{supp}
\DeclareMathOperator{\Const}{Const}

\DeclareMathOperator{\one}{\mathbf{1}}

\DeclareMathOperator{\diam}{{\rm diam}}

\def\cdott{\cdot\,}
\def\eu{\mathrm{e}}

\def\bcZN{\boldsymbol{\mathcal{Z}}^{N}}
\def\bcEN{\boldsymbol{\mathcal{E}}^{(N)}}

\def\emp{\varnothing}
\def\rd{\R{d}}
\def\brd{\B{d}}

\def\rS{\R{S}}

\def\Bx{\B{x}}
\def\By{\B{y}}
\def\Bu{\B{u}}

\def\Bn{\B{n}}

\def\Brho{\boldsymbol{\rho}}
\def\BrhoS{\boldsymbol{\rho}_{\R{S}}}

\def\BPsi{\boldsymbol{\Psi}}

\def\Lam{\Lam}
\def\om{\omega}
\def\Om{\Omega}

\def\BA{\B{A}}
\def\BB{\B{B}}

\def\BH{\B{H}}

\def\BU{\B{U}}
\def\BV{\B{V}}

\def\BDelta{\B{\Delta}}

\def\ball{\R{B}}
\def\bball{\B{B}}

\def\DC{\D{C}}

\def\DP{\D{P}}
\def\DR{\D{R}}
\def\DZ{\D{Z}}
\def\DN{\D{N}}

\def\cE{\C{E}}
\def\cH{\C{H}}
\def\cJ{\C{J}}

\def\cZ{\C{Z}}

\def\RCM{\textsf{(RCM)}\xspace}

\def\fF{\mathfrak{F}}
\def\fS{\mathfrak{S}}
\def\fk{\mathfrak{k}}
\def\fZ{\mathfrak{Z}}

\def\pr#1{\D{P}\left\{\,#1\,\right\}}
\def\esm#1{\D{E}\left\{\,#1\,\right\}}

\def\be{\begin{equation}}
\def\ee{\end{equation}}
\def\ba{\begin{array}{l}}
\def\ea{\end{array}}
\def\bal{\begin{aligned}}
\def\eal{\end{aligned}}

\def\ble{\begin{lemma}}
\def\ele{\end{lemma}}

\def\bthm{\begin{theorem}}
\def\ethm{\end{theorem}}

\def\bco{\begin{corollary}}
\def\eco{\end{corollary}}

\def\bre{\begin{remark}}
\def\ere{\end{remark}}

\def\btm{\begin{theorem}}
\def\etm{\end{theorem}}

\def\bde{\begin{definition}}
\def\ede{\end{definition}}

\begin{document}
\title[A remark  on  charge transfer in multi-particle systems]
{A remark on  charge transfer processes  in multi-particle systems}

\author[V. Chulaevsky]{Victor Chulaevsky}


\address{D\'{e}partement de Math\'{e}matiques\\
Universit\'{e} de Reims, Moulin de la Housse, B.P. 1039\\
51687 Reims Cedex 2, France\\
E-mail: victor.tchoulaevski@univ-reims.fr}

\begin{abstract}
We assess the probability of  resonances between sufficiently distant states
$\Bx=(x_1, \ldots, x_N)$ and $\By=(y_1, \ldots, y_N)$ in a combinatorial graph $\cZ$ serving as
the configuration space of an $N$-particle disordered quantum system. This includes the cases where the transition
$\Bx \rightsquigarrow \By$ "shuffles" the particles in $\Bx$, like the transition $(a,a,b) \rightsquigarrow (a, b, b)$ in a $3$-particle system. In presence of a random external potential $V(\cdot, \omega)$ (Anderson-type models) such pairs of configurations $(\Bx,\By)$ give rise to local (random) Hamiltonians which are strongly coupled, so that eigenvalue (or eigenfunction) correlator bounds are difficult to obtain (cf. \cite{AW09a}, \cite{CS09b}). This  difficulty, which occurs for $N\ge 3$, results in eigenfunction decay bounds  weaker than expected. We show that more efficient bounds, obtained so far only for $2$-particle systems \cite{CS09b}, extend to any $N>2$.

\tblue{In version 4, we extend the techniques from earlier versions (developed for $\cZ = \DZ^d$ [v.1, 19.05.2010])
to more general graphs, adapt them to the fermionic and bosonic systems, and streamline the proofs.
}

\end{abstract}
\maketitle

\section{Introduction. The model} \label{sec:intro}

We study quantum systems in a disordered environment, usually referred to as Anderson-type models, due to the seminal paper by
P. W. Anderson \cite{An58}. For nearly \textit{fifty years} following its publication, the localization phenomena have been studied
in the single-particle approximation, i.e. under the assumption that the interaction between particles subject to the common
disordered (usually understood as "random") external potential is "sufficiently weak" to be neglected in the analysis of the decay
properties of eigenstates of the multi-particle system in question. A detailed discussion of recent developments in the physics of
disordered media is most certainly beyond the scope of this paper; we simply refer to recent papers by Basko--Aleiner--Altshuler
\cite{BAA05} and by Gornyi--Mirlin--Polyakov \cite{GMP05} (the order of citation is merely alphabetical) where it was shown,
in the framework of physical models and methods, that the localization phenomena, firmly established in the non-interacting
multi-particle disordered  quantum systems, persist in presence of non-trivial interactions.

We consider a system of $N\ge 2$ quantum particles in a finite or countable, locally finite graph $\cZ$,
endowed with the canonical graph  distance $\rd(\cdott,\cdot) = \rd_\cZ(\cdott,\cdot)$.
In the first part of the paper,
the particles are considered distinguishable, so the $N$-particle configuration space is the Cartesian product $(\cZ)^N$.
In Sect.~\ref{sec:Fermi.Bose}, we extend our technique to the bosonic and fermionic systems and define the configuration space
properly.

The Hamiltonian is assumed to have the form
\begin{equation}\label{eq:H}
\BH(\om)= \sum_{j=1}^N \left( \Delta^{(j)} + V(x_j, \omega)\right) + \BU,
\end{equation}
where $V:\cZ\times \Omega \to \DR$ is a random field relative to a probability space
$(\Omega, \C{F}, \D{P})$, $\Delta^{(j)}$ is the canonical graph  Laplacian on the $j$-th replica of
the $1$-particle configuration space (graph) $\cZ$, viz.
$$
\Delta^{(j)}  \Psi(x_1, \ldots, x_N) = \sum_{y\in\DZ^d:\, \rd_\cZ(y,x_j)=1}  \Psi(x_1, \ldots, x_j + y,  \ldots, x_N),
$$
and $\BU$ is the multiplication operator by a function $\BU(\Bx)$ which we assume bounded (this assumption can be relaxed).  The symmetry of the function $\BU$ is not required, and we do not assume $\BU$ to be a "short-range" or rapidly decaying interaction. In fact, we focus here on restrictions of $\BH_{V,\BU}$ to finite subsets of the lattice, so that $\BH_{V,\BU}$ may or may not be well-defined on the entire lattice $\DZ^d$: this does not affect our main result.

The assumptions on the random field $V$ are described below, in Section \ref{sec:EV.sep}.

\subsection{Eigenvalue concentration bounds}

We focus here on probabilistic bounds of certain eigenvalue correlators, or eigenvalue concentration bounds, known in the single-particle localization theory as Wegner-type bounds, due to the paper by Wegner \cite{We81}. It would not be an exaggeration to say that this bound is the heart of the MSA. (In a slightly disguised form, it also appears in the framework of the FMM both in its the single-particle and multi-particle version, as the reader can observe in \cites{AM93,AW09a,AW09b}.)
In essence, one may call a Wegner-type bound a (sufficiently explicite and suitable for applications)
probabilistic bound of the form
$$
\pr{ \dist(E, \sigma(\BH_{\Lambda}(\omega))) \le \epsilon} \le f(|\Lambda|, \epsilon),
\eqno({\rm W1})
$$
where $\BH_{\Lambda}(\omega)$ is the restriction of $\BH(\omega)$ on a bounded subset $\Lambda$ with some self-adjoint boundary conditions, and $\sigma(\BH_{\Lambda}(\omega))$ is its spectrum (a finite number of random points, in the case of lattice models).

The role and importance of such bounds can be easily understood: the MSA procedure starts with the analysis of the resolvents
$(\BH_{\Lambda} - E)^{-1}$, so it is vital to know how unlikely it is to have the spectrum of $\BH_\Lambda$ $\epsilon$-close to a given value $E\in\DR$ .

Given any finite ball $\bball_{L}(\Bu) := \{\Bx\in\bcZN\,|\, |\Bx - \Bu| \le L\}$,
we will consider a finite-volume approximation of the Hamiltonian $\BH$
$$
\BH_{\bball_{L}(\Bu)} = \BH \upharpoonright_{\ell^2(\bball_{L}(\Bu))} \text{ with Dirichlet boundary conditions on } \partial \bball_{L}(\Bu).
$$
acting in the finite-dimensional Hilbert space $\ell^2(\bball_{L}(\Bu))$.

In Ref. \cite{CS08}, where the configuration space was $\cZ=\DZ^d$, $d\ge 1$,
the following "two-volume" version of the Wegner bound was established for pairs of two-particle operators $\BH_{\bball_{L}(\Bu)}$, $\BH_{\bball_{L'}(\Bu')}$  such that $L \ge L'$ and
$\dist(\Bu, \Bu') \ge 8L$:
if $\nu$ is the continuity modulus of the marginal distribution of the IID random field $V$, then
$$
\pr{ \dist( \sigma(\BH_{\bball_{L}(\Bu)} ), \sigma( \BH_{\bball_{L'}(\Bu')}) \le \epsilon}\\
\le \, (2L+1)^{2d} (2L'+1)^d \,\nu(2\epsilon).
\eqno({\rm W2})
$$
The proof given in \cite{CS08} is based on a geometrical notion of "separable" pairs of balls, combined with Stollmann's 
lemma \cite{St01}
on
diagonally monotone functions. In \cite{CS07} a similar bound
was proven in the case of IID random field $V$ with analytic marginal distribution.

Unfortunately, starting from $N=3$, additional difficulties appear in the analysis of pairs of spectra $ \sigma(\BH_{\bball_{L}(\Bu)} ), \sigma( \BH_{\bball_{L}(\Bu')})$.  To put it simply,  no a priori lower bound on the distance $\dist(\bball_{L}(\Bu), \bball_{L}(\Bu') )>CL$ between two balls of sidelength $O(L)$ can guarantee the approach of \cite{CS08} to work, no matter how large is the constant $C$. This gives rise to a significantly more sophisticated MPMSA procedure in the general case where  $N\ge 3$. A similar difficulty arises in \cite{AW09a}.

\subsection {The main goal}\label{ssec:motiv}

It is well-known that the FMM, when applicable, leads directly to the proof of the dynamical localization, while it is more natural for applications of the MSA to establish first the spectral localization, via probabilistic bounds of the kernels of resolvents $G_\Lambda(E) = (H_\Lambda-E)^{-1}$ in finite subsets (usually balls) $\Lambda\subset \DZ^d$, and then derive dynamical localization from decay bounds of the resolvents $G_\Lambda(E)$.

In \cites{CS09a,CS09b} a multi-particle adaptation of the MSA was used to prove \textit{spectral} localization (i.e., exponential decay of eigenfunctions) in the strong disorder regime. Aizenman and Warzel \cites{AW09a,AW09b} used the FMM to prove directly \textit{dynamical} localization (hence, spectral localization) in various parameter regions including strong disorder, "extreme" energies and weak interactions.

Despite many differences between these two approaches, similar technical difficulties have been encountered in both cycles of papers. Namely, it turned out to be difficult to prove the decay bounds of eigenfunctions $\Psi_j^{(N)}(x_1, \ldots, x_n)$ of $N$-particle Hamiltonians
in terms of some \textit{norm} $\|\cdot\|$ in $\DR^{Nd}$:
$$
|\Psi_j^{(N)}(x_1, \ldots, x_n;\omega)| \le C_j(\omega) e^{-m \|\Bx\|}.
$$
If the interaction $\BU$ is symmetric (and so is, then, $\BU + \BV$), then it is natural to expect (or to fear ...) "resonances" and  "tunneling" between a point
$\Bx = (x_1, \ldots, x_N)$ and the points $\tau(\Bx) = (x_{\tau(1)}, \ldots, x_{\tau(N)})$ obtained by
permutations $\tau\in \fS_N$. So, it is much more natural in this context to use the symmetrized distance
$$
\dist_\rS(\Bx, \By) := \min_{\tau\in \fS_N} \|\tau(\Bx) - \By\|.
$$
Note also that if the quantum particles are bosons or fermions, then the points $\tau(\Bx)$ should even be treated as identical, or, more precisely,  the spectral problem should be solved in the subspace of symmetric or anti-symmetric functions of variables $x_j$.

However, due to a highly correlated nature of the potential of a multi-particle system, even the above concession did not suffice, and it was easier to use "Hausdorff distance" (see the definition below, in Section \ref{sec:config.WS}) between points $\Bx,\By\in(\DZ^d)^N$. This resulted in weaker decay estimates than expected. (Note that the Hausdorf distance was not used explicitly in \cite{CS09b}.)

Aizenman and Warzel \cite{AW09a} analyzed explicitly the aforementioned technical problem and   pointed out that, physically speaking, it was difficult to rule out the possibility of "tunneling" between points $\Bx$ and $\By$ related  by a "partial charge transfer" process, e.g., between points
$(a, a, b)$ and $(a, b, b)$, $a\ne b$, corresponding to the states:
$$
\begin{array}{l}
\text{ state $\Bx$: 2 particles at the point $a$ and $1$ particle \,\,at $b$}\\
\text{ state $\By$: 1 particle \,\,at the point $a$ and $2$ particles at $b$.}
\end{array}
 $$
Observe that the norm-distance between such states  can be arbitrarily large.

\textit{In the present paper we address this problem and show that resonances between distant states in the configuration space, related by partial charge transfer  processes, are unlikely,  providing  probabilistic estimates for such unlikely situations.}


\subsection{ The main result}\label{ssec:main.res}

Unless otherwise specified, we always work with an arbitrary connected, locally finite graph $(\cZ,\cE)$ with the vertex set $\cZ$ and the
edge set $\cE$; by slight abuse of notations, we often identify the graph with $\cZ$.

We assume that the growth rate of the balls in $\cZ$ is uniformly bounded, viz.
\be\label{eq:growth.general.f}
\sup_{x\in\cZ} | \ball_L(x)| \le f_\cZ(L) < +\infty.
\ee
While a significant part of the techniques and intermediate results in our paper does not require, formally, any upper bound
on the cardinality of the balls, the main application (to the multi-scale analysis) requires a sub-exponential bound on $f_\cZ$.

We will consider in particular the class $\fZ(d,C_d)$ of polynomially growing graphs $\cZ$,
satisfying for some $d,C_d\in(0,+\infty)$
$$
f_\cZ(L) = \sup_{x\in\cZ} | \ball_L(x)| \le C_d L^d, \;\; L\ge 1.
$$
The class of graphs with $f_\cZ(L) \le \Const \eu^{L^\delta}$, $0<\delta<1$, is also suitable for the multi-scale analysis,
but actually the most interesting case after the lattices $\DZ^d$ are the trees and other graphs with exponential growth rate
of balls, and the latter, as is well-known, remain so far beyond the reach of the MSA techniques.

We assume that an IID random field
$V:\cZ\times\Om\to\DR$, relative to a probability space $(\Om,\fF,\DP)$, is defined on $\cZ$.

Introduce the following notations. Given a finite subset  $Q\subset \cZ$, we denote by $\xi_{Q}(\omega)$
the sample mean of the random field $V$ over the $Q$,
$$
 \xi_{Q}(\omega) =  \langle V \rangle_Q  = \frac{1}{| Q |} \sum_{x\in Q} V(x,\omega)
$$
and the "fluctuations" of $V$ relative to the sample mean,
$$
 \eta_x  = V(x,\omega) - \xi_{Q}(\omega), \; x\in Q.
$$
We denote by $\fF_{Q}$ the sigma-algebra generated by $\{\eta_x, \,x\in Q\}$, and by
$F_\xi( \cdot\,| \fF_{Q})$ the conditional probability distribution function of $\xi_Q$ given
$\fF_{Q}$:
$$
 F_\xi(s\,| \fF_{Q}) := \pr{ \xi_Q \le s\,|\, \fF_{Q} }.
$$
For a given $s\in\DR$, $F_\xi(s\,| \fF_{Q})$ is a random variable, determined by the values of $\{\eta_x, x\in Q\}$, but we will often use  inequalities involving it, meaning that these relations hold true for $\D{P}$-a.e. condition.

We will assume that the random field $V$ satisfies the following
condition\footnote{In an earlier version of this manuscript (1005.3387v2, 02.07.2010), we assumed a stronger condition: a uniform continuity of the conditional probability distribution function
$F_\xi(\cdot\,| \fF_{Q})$, i.e., a uniform bound for a.e. condition.}
(RCM = Regularity of the Conditional Mean):
\par\medskip
\RCM:
\textit{ There exist constants $C', C'', A', A'', b', b''\in(0,+\infty)$ such that for any finite subset $Q\subset\cZ$ ,
the conditional probability distribution  function
$F_\xi( \cdot \,| \fF_{Q})$ satisfies for all $s\in(0,1)$
\begin{equation}\label{eq:RCM}
 \;\;
\pr{ \sup_{t\in \DR} \; |F_\xi(t+2s\,| \fF_{Q}) - F_\xi(t\,| \fF_{Q})| \ge C' |Q|^{A'} s^{b'}) }
\le C'' |Q|^{A''} s^{b''}.
\end{equation}
} 

\smallskip

In the particular case of a Gaussian  IID field $V$, e.g.,  with zero mean and unit variance, $\xi_Q$ is a Gaussian random variable with variance ${|Q|}^{-1}$, independent of the "fluctuations" $\eta_x$, so that its probability density is bounded:
$$
p_{\xi_{Q}}(s) = |Q|^{1/2} \, (2\pi)^{-1/2}\, e^{-\frac{{|Q|} s^2}{2 } } \le |Q|^{1/2} \, (2\pi)^{-1/2},
$$
although the $L_\infty$-norm of its probability density grows as $|Q|\to\infty$, and so does the continuity modulus of the distribution function $F_{\xi_{Q}}$.

Formally speaking, the condition \RCM does not refer to the growth rate of balls in the graph $\cZ$,
but the analytic form of the estimate \eqref{eq:RCM} is in fact adapted to the class $\fZ(d,C_d)$.
[ It can be easily extended to the graphs with sub-exponential growth.] The same remark can be made
concerning the formulation of our main result:

\btm\label{thm:W2.general}
Let $V: \cZ\times \Omega \to \DR$ be a random field satisfying
{\rm \RCM}.
Then for any pair of $N$-particle operators $\BH_{\bball_{L'}(\Bu')}$, $\BH_{\bball_{L''}(\Bu'')}$,
$0 \le L', L'' \le L$, satisfying $\BrhoS(\Bu', \Bu'') > (4N-2)L$, and any $s>0$ the following bound holds:
\begin{equation}\label{eq:main.bound}
\pr{ \dist(\sigma(\BH_{\bball_{L'}(\Bu')}), \sigma(\BH_{\bball_{L''}(\Bu'')})) \le s }
= h_L(s)
\end{equation}
with
\be\label{eq:def.h.L}
h_L(s) := |\bball_{L''}(\Bx)| \cdot |\bball_{L''}(\By)|\, C'L^{A'} s^{b'} + C''L^{A''} s^{b''}.
\ee
\etm

In general, the conditional distribution function $F_\xi(\cdot\,|\fF_{Q})$ is not necessarily uniformly continuous, let alone H\"{o}lder-continuous. Moreover, the following simple example shows that for some conditions the distribution of the sample mean can be extremely singular.

\begin{example}\label{ex:unif}
Let $v_1(\omega), v_2(\omega)$ be two independent random variables uniformly distributed in $[0,1]$. Set
$\xi = (v_1+v_2)/2$, $\eta = (v_1-v_2)/2$. Conditioning on $\eta\ge 0$ induces a uniform probability distribution on the segment $I(\eta)=\{ (t+2\eta,t), t \in(0, 1-2\eta)\}$ of length $|I(\eta)| = 1-2\eta$, with constant probability density $(1-2\eta)^{-1}$, if $\eta< 1/2$. Obviously, these distributions are not uniformly continuous. Moreover, for $\eta=1/2$, $\xi$ takes a single value: $\xi = 1/2$, so that its conditional distribution is no longer continuous. Observe, however, that "singular" conditions have probability zero, and conditions which give rise to large conditional density of $\xi$ have small probability.
\end{example}


\section{Distinguishable particle configurations and weak separability } \label{sec:config.WS}

\subsection{Basic definitions}

Given a connected graph $(\cZ,\cE)$ and an integer $N\ge 2$, introduce the product graph
$(\bcZN, \bcEN)$ with the vertex set $\bcZN \equiv (\cZ)^N$ and the edge set defined as follows:
with $\Bx=(x_1,\ldots,x_N)$, $\By=(y_1,\ldots,y_N)\in\bcZN$,
$$
(\Bx,\By)\in\bcEN \Leftrightarrow \sum_{j=1}^N \rd_\cZ(x_j,y_j) = 1.
$$
Observe that this definition gives the conventional graph structure on the lattice $\bcZN$ considered
as the product $(\DZ^d)^N$, $d\ge 1$, $N\ge 2$.

Intervals of integer values will often appear in our formulae, and it is convenient to use a standard notation $[[a,b]] := [a,b]\cap \DZ$.

We identify $N$-tuplets $\Bx\in\bcZN$ with  configurations of $N$ distinguishable particles in $\cZ$:
$\Bx \equiv (x_1, \ldots, x_N) \in \DZ^d \times \cdots \times \DZ^d$.

Pictorially, $\By$ is a nearest graph neighbor of $\Bx$ in $\bcZN$ if it is obtained from $\Bx$ by moving exactly
one particle $x_j$ to an adjacent vertex $y_j\in\cZ$, and vice versa.

The graph structure defines in $\bcZN$ the canonical graph distance $\brd_{\bcZN}$: for $\Bx \ne \By$,
$\brd_{\bcZN}(\Bx, \By)$ is the length of the shortest path $\Bx \rightsquigarrow \By$ over the edges of $\bcZN$.
With $\cZ$ fixed, we will often drop the subscript and simply write $\brd(\Bx, \By)$.

One can also define the max-distance on $\bcZN$:
$$
\Brho(\Bx,\By) := \max_{j\in[[1,N]]} \rd(x_j, y_j).
$$

It turns out that the graph distance $\brd$ and the max-distance $\Brho$ are not well-adapted to
the analysis of quantum resonances in the multiparticle systems,
so we introduce the symmetrized max-distance by
$$
\BrhoS(\Bx,\By) := \min_{ \pi\in\fS_N} \Brho\big( \pi(\Bx), \By) )
\equiv \min_{ \pi\in\fS_N} \Brho\big( \pi(\By), \Bx) ) .
$$
The importance of the symmetrized sistance can be explained as follows. First, assume that the interaction $\BU$ is permutation
symmetric; the non-interacting Hamiltonian
$$
\BH^{\rm{ni}} = \sum_{j=1}^N \left( H_{0;j} + gV(x_j;\om) \right)
$$
being always permutation symmetric, the full Hamiltonian $\BH = \BH^{\rm{ni}} + \BU$ then also is permutation invariant.
Therefore, if $\Bx = \pi(\By)$ for some $\pi \ne {\rm Id}$, the local Hamiltonians
$\BH_{\bball_L(\Bx)}$ and $\BH_{\bball_L(\By)}$ have identical spectra, no matter how far apart are the centers $\Bx$ and $\By$.
This renders impossible any kind of EVC bounds in the course of the localization analysis. In fact, this difficulty is still
present even if $\BU$ is not permutation symmetric, for the probabilistic analysis of eigenvalue concentration relies upon the
external random potential $\BV(\Bx;\om) = \sum_j V(x_j;\om)$, which still is permutation symmetric.

\begin{definition} \label{def:subconfig}
Let $\Bx\in\bcZN$ and consider  a  subset
$\cJ \subset[[1,N]]$ with $1\le |\cJ|=n < N$. A \textbf{subconfiguration} of $\Bx$ associated with  $\cJ$ is the pair $(\Bx', \cJ)$ where the vector
$\Bx' \in \bcZN$ has the components $x'_i = x_{j_i}$, $i\in[[1, n]]$. Such a subconfiguration will be denoted as $\Bx_{\cJ}$. The complement of a
subconfiguration $\Bx_{\cJ}$ is the subconfiguration $\Bx_{\cJ^c}$  associated with the complementary index subset $\cJ^c := [[1,N]]\setminus \cJ$.
\end{definition}

By a slight abuse of notations, we will  identify a subconfiguration
$\Bx_\cJ = (\Bx', \cJ)$ with the vector $\Bx'$. With $\cJ$ clearly identified (this will always be the case in our arguments), it should not lead to any ambiguity, while making notations simpler.

\begin{definition} \label{def:sep.balls}
(a) Let $N\ge 2$ and consider the set of all $N$-particle configurations $\bcZN$. For each
$j\in[[1,N]]$  the coordinate projection $\Pi_j:\bcZN\to \cZ$ onto the coordinate space of the $j$-th particle is the mapping
$$
\Pi_j: (x_1, \ldots, x_N) \mapsto x_j.
$$
(b) The support $\Pi \Bx$  of a configuration $\Bx\in\bcZN$, $n\ge 1$, is the set
$$
\Pi \Bx := \cup_{j=1}^n \Pi_j \Bx = \{x_1, \ldots, x_N \}.
$$
Similarly, the support of a subconfiguration $\Bx_\cJ$  is defined by
$
\Pi \Bx_\cJ := \cup_{j\in \cJ}^n \Pi_j \Bx.
$
\par
\smallskip
\noindent
(c) Given a subset  $\cJ\subset [[1,N]]$ with $|\cJ|=n$, the projection
$\Pi_\cJ: \,\bcZN \to \cZ$ is defined as follows:
$$
\Pi_\cJ \Bx =
\begin{cases}
\Pi \Bx_\cJ, & \text{ if } \cJ \ne \varnothing \\
\varnothing,                & \text{ otherwise}.
\end{cases}
$$
Finally, for each subset $\bball\subset \bcZN$  its support 
$\Pi \bball$ is defined by
$$
\Pi \bball := \bigcup_{j=1}^N \Pi_j \bball \subset \cZ.
$$
\end{definition}

We will not associate with the  empty subconfiguration $\Bx_\varnothing$ any object other than its support $\Pi \Bx_\varnothing = \varnothing \subset \cZ^d$, so the above definitions and notations suffice for our purposes.

Particle configurations being associated with  point subsets of $\cZ$, one can introduce the distance between two arbitrary configurations $\Bx'\in \DZ^{N'd}$,
$\Bx''\in \DZ^{N''d}$, $N', N'' \ge 1$, as the distance between the respective subsets of $\cZ$, induced by the max-norm:
$$
\begin{array}{c}
\Brho\left(\Bx', \Bx'' \right)
\equiv \Brho\left( \{x'_1, \ldots, x'_{N'} \}, \{x''_1, \ldots, x''_{N''} \} \right) \\
\quad = \displaystyle \min_{i\in[[1, N']]} \; \min_{j\in[[1, N'']]}
\|x'_i - x''_j\|.
\end{array}
$$

\subsection{Weakly separable balls}\label{ssec:def.PCT}
\par
\medskip

\begin{definition}
A  ball  $\bball_L(\Bx)$ is weakly separable from $\bball_L(\By)$ if
there exists a bounded subset $Q\subset \cZ$ in the 1-particle configuration space,
and  subsets $\cJ_1, \cJ_2\subset [[1,N]]$  such that
$|\cJ_1| > |\cJ_2|$ (possibly, with $\cJ_2=\varnothing$) and
\begin{equation}\label{eq:cond.WS}
\begin{array}{l}
 \big( \Pi_{\cJ_1} \bball_L(\Bx) \cup \Pi_{\cJ_2} \bball_L(\By) \big) \;  \subseteq Q,\\
\big( \Pi_{\cJ^c_1} \bball_L(\Bx) \cup \Pi_{\cJ^c_2} \bball_L(\By) \big) \cap Q = \varnothing.
\end{array}
\end{equation}
A  pair of balls $(\bball_L(\Bx), \bball_L(\By))$ is weakly separable if at least one of the balls is weakly separable from the other.
\end{definition}

The physical meaning of the weak separability is that in a certain region of the one-particle configuration space the presence of particles from configuration $\Bx$ is more important than that of the particles from $\By$. As a result, some local fluctuations of the random potential $V(\cdot;\omega)$ have a stronger influence on $\Bx$ than on $\By$.

The conditions \eqref{eq:cond.WS} shows, as does the application of the notion of weak separation to the proof of Lemma \ref{lem:main.lemma}
below, that one can take in \eqref{eq:cond.WS} the minimal set $Q$, i.e.,
$Q = \Pi_{\cJ_1} \bball_L(\Bx) \cup \Pi_{\cJ_2}$. We keep a more general form \eqref{eq:cond.WS}
which may prove useful in some applications.

\ble\label{lem:dist.are.WS}
Balls $\bball_L(\Bx), \bball_L(\By)$ with  $\BrhoS(\Bx, \By) >  (4N-2)L$ are weakly separable.
\ele

\proof

Let $\B{\Gamma}(\Bx)=\{ \Gamma_1, \ldots, \Gamma_M \}$, $1\le M \le N$,
be the collection of clusters of the $1$-particle balls $\ball_{2L}(x_j)$,
i.e., of the minimal pairwise disjoint connected components of the union of the balls $\ball_{2L}(x_j)$,
so that
$[[1,N]]= \sqcup_{1\le i\le M} \cJ_i$, and $\Gamma_i = \cup_{j\in\cJ_i} \ball_{2L}(x_j)$.

Denote $Q_i =  \cup_{j\in\cJ_i} \ball_{L}(x_j)$.
Since the clusters are connected, we have
$$
\diam(Q_i) \le \sum_{j\in\cJ_i} \diam\big( \ball_{2L}(x_j) \big) \le 4 N L .
$$
Observe that $\ball_{2L}(x_j)\cap\ball_{2L}(x_k)\ne \emp \Leftrightarrow \rd(x_j,x_k)\le 4L$, thus
$$
\forall\, i=1, \ldots, M\quad \diam \{ x_j, \; j\in\cJ_i\} \le 4L(N-1),
$$
and if $y\in \Gamma_i$, then $\min_{j\in\cJ_i} \rd(y,x_j) \le 2L$, yielding
\be\label{eq:dist.y.all.xj.Gamma_j}
\bal
\max_{j\in\cJ_i} \rd(y,x_j) & \le 2L + \diam \{ x_j, \; j\in\cJ_i\} \le 2L + 4L(N-1)
\\
& = (4N-2)L.
\eal
\ee
Introduce the occupation numbers of the sets $\Gamma_i$ for  configurations $\Bx$ and $\By$:
\begin{align*}
n_i(\Bx) = \card\left(  \Pi \Bx \cap \Gamma_i \right), \; i\in[[1,M]],\\
n_i(\By) = \card\left(  \Pi \By \cap \Gamma_i \right), \; i\in[[1,M]].
\end{align*}
It follows immediately from the definition of $\Gamma_i$ that $n_i(\Bx) = |\cJ_i|$, $i=1, \ldots, M$,
thus $\sum_i n_i(\Bx) = N$, while for the configuration $\By$ we only have, in general, that $0 \le \sum_i n_i(\By) \le N$.

There can be two possible situations:

\begin{enumerate}[(I)]
  \item  For all $i\in[[1,M]]$ we have
  $n_i(\Bx) = n_i(\By)$. Then by \eqref{eq:dist.y.all.xj.Gamma_j},
  there exists a permutation $\tau\in\mathfrak{S}_N$ such that for all $j\in[[1,N]]$,
$$
\|x_{\tau(j)} - y_j\| \le (4N-2)L,
$$
yielding
$$
\rd_S(\Bx, \By) \le \|\tau(\Bx) - \By \| = \max_{1\le j \le N} \|x_{\tau(j)} - y_j \|
\le 4NL - 2L < 4NL.
$$
If $\rd_S(\Bx, \By)> (4N-2)L$, then the occupation numbers $n_i(\Bx), n_i(\By) $ cannot be all identical,
so this situation is impossible under the hypotheses of the lemma.

  \item For some $i\in[[1,M]]$, $n_i(\Bx) \ne n_i(\By)$. By the definition of $Q_i$,
  it contains $|\Gamma_i|\ge 1$ particles of the configuration $\Bx$, so that $n_i(\Bx)\ge 1$ for all $i\in[[1,M]]$. Observe that
\begin{equation}\label{eq:occup.num}
\sum_{i=1} ( n_i(\Bx) - n_i(\By)) = N - \sum_{i=1} n_i(\By) \ge 0.
\end{equation}
Since not all quantities $n_i(\Bx) - n_i(\By)$ vanish, there exists some
$j_\circ\in [[1,M]]$ such that $n_{j_\circ}(\Bx) - n_{j_\circ}(\By)>0$ , otherwise the LHS of \eqref{eq:occup.num} would be negative.

  Now setting $Q = Q_{j_\circ}$, we see that the conditions \eqref{eq:cond.WS} are fulfilled.
\end{enumerate}
\qedhere

\section{Eigenvalue concentration bound for distant balls } \label{sec:EV.sep}

\subsection{Bounds for weakly separable balls}\label{ssec:W2.WS}

\ble\label{lem:main.lemma}

Let $V: \cZ\times \Omega \to \DR$ be a random field  satisfying the condition
\RCM. Let $\Bx,\By\in \cZ$ be two configurations such that the balls
$\bball_{L}(\Bx)$, $\bball_{L}(\By)$ are weakly separable. Consider operators $\BH_{\bball_{L'}(\By)}(\omega)$, $\BH_{\bball_{L''}(\By)}(\omega)$, with $L', L''\le L$.
Then for any $s>0$ the following bound holds for the  spectra
$\Sigma_\Bx=\sigma(\BH_{\bball_{L'}(\Bx)})$, $\Sigma_\Bx=\sigma(\BH_{\bball_{L''}(\By)})$
of these operators:
$$
\bal
\pr{ \dist(\Sigma_\Bx,\Sigma_\By)) \le s }
%
\le |\bball_{L}(\Bu')| \, |\bball_{L}(\Bu'')|\, C'L^{A'} (2s)^{b'}
+ C''L^{A''} (2s)^{b''}.
\eal
$$
\ele

\proof
Let $Q$ be a ball satisfying the conditions \eqref{eq:cond.WS}
for some $\cJ_1, \cJ_2 \subset [[1,N]]$ with  $|\cJ_1|=n_1 > n_2 =|\cJ_2|$.
Introduce the sample mean $\xi=\xi_{Q}$ of $V$ over $Q$ and the fluctuations
$\{\eta_x, \, x\in Q \}$  defined as in Section \ref{ssec:main.res}.

The operators $\BH_{\bball_{L'}(\Bx)}(\omega)$,  $\BH_{\bball_{L''}(\By)}(\omega)$ read as follows:
\begin{equation}\label{eq:Ham.decomp}
\begin{array}{l}
\BH_{\bball_{L'}(\Bx)}(\omega) = n_1 \xi(\omega) \, \one + \BA(\omega), \\
\BH_{\bball_{L''}(\By)}(\omega) = n_2 \xi(\omega) \,\one + \BB(\omega)
\end{array}
\end{equation}
where operators $\BA(\omega)$ and $\BB(\omega)$ are $\fF_{Q}$-measurable. Let
$
\{ \lambda_1, \ldots, \lambda_{M'}\}, \;
M' = \,\,|\bball_{L'}(\Bx)|,
$
and
$
\{ \mu_1, \ldots, \mu_{M''}\},
M'' = |\bball_{L''}(\By)|,
$
be the sets of eigenvalues of $\BH_{\bball_{L'}(\Bx)}$ and of $\BH_{\bball_{L''}(\By)})$,
counted with multiplicities.
Owing to \eqref{eq:Ham.decomp}, these eigenvalues can be represented as follows:
$$
\begin{array}{l}
\lambda_j(\omega) = n_1\xi(\omega) + \lambda_j^{(0)}(\omega),
\quad
\mu_j(\omega) = n_2\xi(\omega) + \mu_j^{(0)}(\omega),
\end{array}
$$
where the random variables
$\lambda_j^{(0)}(\omega)$ and $\mu_j^{(0)}(\omega)$ are $\fF_{Q}$-measurable. Therefore,
$$
\lambda_i(\omega) - \mu_j(\omega) =  (n_1-n_2)\xi(\omega) + (\lambda_j^{(0)}(\omega) -  \mu_j^{(0)}(\omega)),
$$
with $n_1-n_2 \ge 1$, owing to our assumption.
Further, we can write
$$
\bal
\pr{ \dist(\Sigma_\Bx, \Sigma_\By)) \le s }
&= \pr{ \exists\, i,j:\, |\lambda_i - \mu_j| \le s }
\\
%
& \le \sum_{i=1}^{M'} \sum_{j=1}^{M''}
     \esm{ \pr{ |\lambda_i - \mu_j| \le s \,| \fF_{Q}}}.
\eal
$$
Note that for all $i$ and $j$ we have, with $n_1 > n_2$ and $(n_1 - n_2)^{-1} \le 1$,
$$
\bal
\pr{ |\lambda_i - \mu_j| \le s \,|\, \fF_{Q}}
& = \pr{ |(n_1 - n_2)\xi + \lambda_i^{(0)} - \mu_j^{(0)}| \le s \,| \fF_{Q}}
\\
&\le \displaystyle \sup_{t\in\DR} \pr{ |\xi - t| \le 2|n_1 - n_2|^{-1} s\,|\, \fF_{Q} }
\\
& \le \sup_{t\in\DR} \pr{ |\xi - t| \le s \,|\, \fF_{Q} }
\\
& \le \sup_{t\in\DR} \big( F_\xi( t + 2s \,|\, \fF_{Q}) - F_\xi( t \,|\, \fF_{Q}) \big)
\eal
$$
Introduce the event
$$
\cE_L = \Bigl\{\; \sup_{t\in\DR} \;
\big|F_\xi(t+2s\,| \fF_{Q}) - F_\xi(t\,| \fF_{Q})\big|\ge C'L^{A'} s^{b'} \Bigr\} .
$$
By the hypothesis \RCM
(cf. \eqref{eq:RCM}), $\pr{\cE_L} \le C''L^{A''} s^{b''}\}$.
Therefore,
$$
\bal
\pr{ \dist(\Sigma_\Bx, \Sigma_\By)) \le s }
&= \esm{ \pr{  \dist(\Sigma_\Bx, \Sigma_\By) \le s  \,| \fF_{Q}}}
\\
&\le \esm{ \one_{\cE^c_L} \pr{ \dist(\Sigma_\Bx, \Sigma_\By) \le s \,| \fF_{Q}}}
+ \pr{\cE_L}
\\
&\le |\bball_{L''}(\Bx)| \cdot |\bball_{L''}(\By)|\, C'L^{A'} s^{b'}
+ C''L^{A''} s^{b''}
= h_L(s),
\eal
$$
with $h_L$ defined in \eqref{eq:def.h.L}.
\qed

\subsection{Proof of the main result}\label{ssec:proof.main.thm}

\par
\medskip
$\,$
\par
\medskip

By the hypothesis of Theorem \ref{thm:W2.general}, we have $\BrhoS(\Bx, \By) > (4N-2)L$;
therefore, by Lemma \ref{lem:dist.are.WS}, balls $\bball_{L'}(\Bx)$ and $\bball_{L''}(\By)$ are weakly separable. Now the assertion of the theorem follows from Lemma \ref{lem:main.lemma}.
\qedhere

\section{EVC bounds for the  particle systems with quantum symmetry}
\label{sec:Fermi.Bose}

The main EVC bound, established for the Hamiltonian $\BH(\om)$ in the entire Hilbert space
$\ell^2(\bcZN)$, implies a lower bound on inter-spectral spacings for the restrictions of $\BH(\om)$ to the
subspaces of symmetric and of anti-symmetric functions, i.e., for the bosonic and fermionic $N$-particle Hamiltonians.
For the convenience of further references, we introduce below required notations and objects.

An technically convenient alternative to restricting $\BH$ to a subspace of $(\pm)$-symmetric functions $\BPsi:(\cZ)^N\to\DC$
consists in representing these subspaces as the spaces of square-summable functions on a reduced graph. Such a reduction
is the most simple in the case where $\cZ = \DZ^1$. Indeed, in this case:
\begin{itemize}
  \item the subspace of anti-symmetric functions $\BPsi:(\DZ)^N\to\DC$ in the Hilbert space $\cH^{(N)}=\ell^2\big(\DZ^N\big)$
  is canonically isomorphic to the space $\ell^2\big(\DZ^N_{>}\big)$ with
$$
\DZ^N_{>} := \{\Bx=(x_1, \ldots, x_N)\in\DZ^N:\; x_N > x_{N-1} > \cdots > x_1\}.
$$
  Up to a constant factor of $(N!)^{1/2}$, the canonical isomorphism is given by the restriction operator
  $\BPsi \mapsto \big(\one_{\DZ^N_\ge}\BPsi \big) \upharpoonright_{\DZ^N_\ge}$.

  \item Similarly,  the subspace of symmetric functions $\BPsi:(\DZ)^N\to\DC$ in $\cH^{(N)}=\ell^2\big(\DZ^N\big)$
  is canonically isomorphic to a weighted space $\ell^2\big(\DZ^N_{\ge}, \fk\big)$ with
$$
\DZ^N_{\ge} := \{\Bx=(x_1, \ldots, x_N)\in\DZ^N:\; x_N \ge  x_{N-1} \ge \cdots \ge x_1\}
$$
and the combinatorial weight $\fk$ determined by the multiplicities of the duplicate components (if any are present)
in a configuration $(x_1, \ldots, x_N)$.
We omit the explicit formula, for it is more natural to start with a self-consistent representation
of a system of indistinguishable particles, without referring to an artificial numeration of their positions.
\end{itemize}

In a more general case, there are alternative constructions of the reduced graph.

\subsection{Fermionic graph}

The standard construction of a symmetric power of an arbitrary locally finite graph $\cZ$ is most suitable for the fermionic
systems. Recall that, by definition, for $N\ge 2$, the $N$-th symmetric power of the graph $(\cZ,\cE)$ is the graph with the vertex set
formed by the $N$-tuplets $\Bx=(x_1, \ldots, x_N)$ with $x_j\in\cZ$, $j\in[[1,N]]$, and
$\card( \{x_1, \ldots, x_N\} = N$, i.e., without duplicate positions in the graph $\cZ$. Clearly, this is a subset
of the edge set of the product graph $\bcZN$ considered before. The edges are those inherited from $\bcEN$:
$(\Bx,\By)$ is an edge iff $\sum_j \rd_\cZ(x_j,y_j)=1$, or, equivalently, if for some pairwise distinct $z_2, \ldots, z_N$,
$$
\Bx = \{ x, z_2, \ldots, z_N\}, \; \By = \{ y, z_2, \ldots, z_N\}, \;\; \card\{x, y, z_2, \ldots, z_N\} = N+2.
$$
As before, these relations can be interpreted in the following way: the configuration of $N$ indistinguishable
particles $\By$ is obtained by moving exactly one particle from the configuration $\Bx$ (without duplicate positions)
to one of its nearest neighbors in $\cZ$.

\subsection{Representation by the occupation numbers}

An alternative construction, which we present first in the fermionic case, is easily adapted to the bosonic
systems.

A configuration of $N$ indistinguishable particles, $N\ge 1$, is uniquely determined by a ``$\DN$-decorated''
subset\footnote{In other words, we consider formal finite linear combinations of vertices from $\cZ$
with non-negative integer coefficients.}
of $\cZ$
formed by all particle positions along with their respective multiplicities. Specifically, introduce the functions
$\Bn_\Bx: \cZ\to \DN$, associated with the indistinguishable particle configurations $\Bx$, with the value
$\Bn_\Bx(u)$ interpreted as the number of particles at $u\in\cZ$ from the configuration $\Bx$. We require that
$$
\sum_{u\in\cZ} \Bn_\Bx(u) = N.
$$

In the fermionic case, we require in addition that
$$
\Bn_\Bx:\cZ \to \{0,1\},
$$
which is a tantamount to assuming the particle positions to be pairwise distinct.
To define the required graph structure in the set of the  occupation number functions, call a pair $(\Bn', \Bn'')$
an edge iff
$$
\sum_{u\in\cZ} |\Bn'(u) - \Bn''(u)| = 2, \;\; \diam \big(\supp (\Bn' -\Bn'') \big) = 1.
$$
In other words, $\supp (\Bn' -\Bn'') = \{x,y\}$, with $\rd(x,y)=1$, so $(x,y)\in\cE$ is an edge in $\cZ$.

The vertex set of the $N$-fermionic graph over $\cZ$ will be denoted by $\bcZN_{-}$.

\subsection{Bosonic graph}

Now we consider all functions $\Bn_\Bx: \cZ\to [[0,N]]$ obeying
$$
\sum_{u\in\cZ} \Bn_\Bx(u) = N.
$$
This allows for duplicate positions; for example, one can have $\Bn_\Bx = N \one_{x}$, interpreted as the configuration
$\Bx=(x, x, \ldots, x)$, with $N$ particles occupying the same position $x\in\cZ$.
The edges are still defined by the constraint
$$
\sum_{u\in\cZ} |\Bn'(u) - \Bn''(u)| = 2, \;\; \diam \big(\supp (\Bn' -\Bn'') \big) = 1.
$$
meaning, as before, that exactly one particle from $\Bx$ is moved to one of its neighboring positions.
For example, $\Bx=(a,a,b)$ and $\By=(a,b,b)$ with $\rd(a,b)=1$ form an edge:
$$
\Bn_\Bx = 2\one_{a} +\one_{b}, \;\; \Bn_\By = \one_{a} +2\one_{b},
$$
so $\Bn_\Bx -\Bn_\By = \one_a - \one_b$, and we have
$$
\supp (\Bn_\Bx -\Bn_\By) = \{a,b\}, \;\; \diam \supp (\Bn_\Bx -\Bn_\By) = 1.
$$

The vertex set of the $N$-bosonic graph over $\cZ$ will be denoted by $\bcZN_{+}$.

\subsection{Self-consistent representation of bosonic/fermonic Hamiltonians}

The following formula provides an equivalent form of the restriction of the $N$-particle Hamiltonian
to the subspace of symmetric ($+$) or anti-symmetric ($-$) functions in $\ell^2\big(\cZ)^N\big)$, without
using any specific order/numeration of the particle positions:
$$
\BH_{\pm}^{(N)}(\om) = \BDelta_{\bcZN_{\pm}} + g \sum_{x\in\Bx} V(x;\om) + \sum_{\substack{x,y\in\Bx\\x\ne y}} U(|x-y|).
$$

\subsection{Balls in the fermionic graph}

We focus now on the fermionic case, which corresponds to the physical model of $N$ electrons
(which are fermions) in the tight binding approximation.

The choice of the metric, defining the notion of a ball in the $N$-particle configuration space,
depends upon the analytic techniques used on the localization analysis, and it is of course not unique.
We consider the case where the max-distance is used; then the balls, like in the case where
$\cZ = \DZ^d$, can be described as polydisks, or suitable subsets thereof, taking into account the quantum symmetry.
Specifically,
$$
\bball_L(\Bx) = \{\By\in\bcZN_-:\, \Brho(\Bx,\By) \le L\}
$$
There is no need here to make use of the subscript "$\rS$" in $\Brho(\cdot\,,\cdot)$, as we did in the first part of the paper,
since the symmetry is now encoded in the very construction of the fermionic graph $\bcZN_-$.

The following example
clearly explains the difference between the
ball $\bball_L(\Bx)$ and the Cartesian product of its single-particle projections $\ball_L(x)\subset\cZ$,
$x\in\Bx$.

\vskip2mm
\noindent
\textbf{Example.} Let $N=2$, $\cZ = \DZ^1$, $L=1$, $\Bx = \{0,1\}$ and $\By = \{0,3\}$.
Then
$$
\bal
\ball_1(\Bx) &= \big\{\, \{1,2\}, \{0,1\}, \{0,2\}, \{-1,0\}, \{-1,1\}, \{-1,2\} \, \big\}
\\
& \subsetneq [-1,1] \times [0,2] \equiv \ball_1(0) \times \ball_1(1),
\eal
$$
while
$$
\ball_1(\By)= [-1,1] \times [2,4] \equiv \ball_1(0) \times \ball_1(3).
$$

\subsection{The EVC bound}

\btm\label{thm:W2.symmetric.graph}
Let $V: \cZ\times \Omega \to \DR$ be a random field satisfying
{\rm \RCM}.
Then for any pair of $N$-particle fermionic Hamiltonians $\BH^{(N,-)}_{\bball_{L'}(\Bu')}$, $\BH^{(N,-)}_{\bball_{L''}(\Bu'')}$,
$0 \le L', L'' \le L$, satisfying $\BrhoS(\Bu', \Bu'') > (4N-2)L$, and any $s>0$ the following bound holds:
\begin{equation}\label{eq:main.bound.sym}
\pr{ \dist(\sigma(\BH^{(N,-)}_{\bball_{L'}(\Bu')}), \sigma(\BH^{(N,-)}_{\bball_{L''}(\Bu'')})) \le s } = h_L(s)
\end{equation}
with
\be\label{eq:def.h.L.sym}
h_L(s) := |\bball_{L'}(\Bx)| \cdot |\bball_{L''}(\By)|\, C'L^{A'} s^{b'} + C''L^{A''} s^{b''}.
\ee
The same bound holds true for the pair of $N$-particle bosonic Hamiltonians $\BH^{(N,+)}_{\bball_{L'}(\Bu')}$,
$\BH^{(N,+)}_{\bball_{L''}(\Bu'')}$, except that the volumes of the bosonic balls
$\bball_{L'}(\Bx)$, $\bball_{L''}(\By)$ may be different from their fermionic counterparts.
\etm

\proof The claim follows from Theorem \ref{thm:W2.general}; it can also be proved directly for the Hamiltonians
in the fermionic/bosonic graphs $\bcZN_{\pm}$, repeating the proof of Theorem \ref{thm:W2.general} almost
verbatim, with minor notational adaptations.
\qedhere






\begin{bibdiv}
\begin{biblist}


\bib{AM93}{article}{
   author={Aizenman, M.},
   author={Molchanov, S. A.},
   title={Localization at large disorder and at extreme energies: an
   elementary derivation},
   journal={Comm. Math. Phys.},
   volume={157},
   date={1993},
   number={2},
   pages={245--278},
}


\bib{AW09a}{article}{
   author={A{i}zenman, M.},
   author={Warzel, S.},
   title={Localization bounds for multi-particle systems},
   journal={Comm. Math. Phys.},
   volume={290},
   date={2009},
   number={3},
   pages={903--934},
}

\bib{AW09b}{article}{
   author={Aizenman, M.},
   author={Warzel, S.},
   title={Complete Dynamical Localization in Disordered
   Quantum Multi-Particle Systems},
   status={arXiv:math-ph/0909:5434v2 (2009)},
   date={2009},
}


\bib{An58}{article}{
   author={Anderson, Ph.},
   title={Absence of diffusion in certain random lattices},
   journal={Phys. Review},
   volume={109},
   date={1958},
   pages={1492--1505},
}

\bib{BAA05}{misc}{
   author={Basko, D. M. },
   author={Aleiner, I. L.},
   author={Altshuler, B. L.},
   title={Metal-insulator transition in a weakly
    interacting many-electron system with
    localized single-particle states},
   status={arXiv:cond-mat/0506617v2},
   date={2005},
}




\bib{CS07}{misc}{
   author={Chulaevsky, V.},
   author={Suhov, Y.},
   title={Anderson localisation for an interacting two-particle quantum system on $\DZ$},
   status={arXiv:math-ph/0705.0657},
   date={2007},
}

\bib{CS08}{article}{
   author={C{h}ulaevsky, V.},
   author={Suhov, Y.},
   title={Wegner bounds for a two-particle tight binding model},
   journal={Comm. Math. Phys.},
   volume={283},
   date={2008},
   number={2},
   pages={479--489},
}
\bib{CS09a}{article}{
   author={Ch{u}laevsky, V.},
   author={Suhov, Y.},
   title={Eigenfunctions in a two-particle Anderson tight binding model},
   journal={Comm. Math. Phys.},
   volume={289},
   date={2009},
   number={2},
   pages={701--723},
}
\bib{CS09b}{article}{
   author={Chu{l}aevsky, V.},
   author={Suhov, Y.},
   title={Multi-particle Anderson localisation: induction on the number of
   particles},
   journal={Math. Phys. Anal. Geom.},
   volume={12},
   date={2009},
   number={2},
   pages={117--139},
}







\bib{GMP05}{misc}{
   author={Gornyi, I. V. },
   author={Mirlin, A. D.},
   author={Polyakov, D. G.},
   title={Interacting electrons in disordered wires: Anderson localization and low-T transport},
   status={arXiv:cond-mat/0506411v1},
   date={2005},
}

\bib{St01}{book}{
   author={Stollmann, P.},
   title={Caught by disorder},
   note={Bound states in random media},
   series={Progress in Mathematical Physics},
   volume={20},
   publisher={Birkh\"auser Inc.},
   place={Boston},
   date={2001},
}


\bib{We81}{article}{
   author={Wegner, F.},
   title={Bounds on the Density of States in Disordered Systems},
   journal={Z. Phys. B, Condensed Matter},
   volume={44},
   date={1981},
   pages={9--15},
}


\end{biblist}
\end{bibdiv}
\end{document}